# Low-frequency one-electrode discharge in long tubes at low gas pressure

Shishpanov A.I., Bazhin P.S., Ivanov D.O. and Meschanov A.V.

*Saint Petersburg State University, Department of optics*

sispanov@mail.ru

The specific form of one electrode capacitive discharge was studied in long tubes filled with high purity neon or argon at pressure 1-4 Torr. The main feature of the discharge is the low rate (less than 10 kHz) of the voltage pulses of given polarity which are applied to only one electrode of the tube, while another one remains free or missing. This type of the discharge was named one-electrode discharge (OED), and seems not to be described in previous studies. The discharge is observed as a glowing plasma column which occupies either tube or its part depending on actual voltage amplitude and on its rate. Current-volt characteristics, ignition thresholds and the OED length changing patterns demonstrate features unknown for RF discharges. It was found that the plasma generation mechanism in OED consists in formation and traveling of the set of ionization waves (IW). As a result, the discharge glow as well as its current can be presented as a set of pulses with duration equal to the wave propagation time (~ 1μs), that appear with the voltage frequency. The pulse has a specific form reflecting the IW structure which is essentially the front of high electrical potential and of plasma channel linking it with the electrode. It was shown that the wave motion is characterized by a pronounced attenuation the patterns of which were investigated by the time-position diagrams method. The attenuation specifies the length of the occupied plasma area as well as other OED parameters. Based on the supposition that the attenuation is the consequence of the IW front potential decrease caused by exponential falling of electric field strength in the plasma channel, the simplified kinematic model of the wave propagation was described. This model allows to estimate the electric field in different OED points as well as average electron concentration via the current measurements; the typical values of the above parameters are 5V/cm and $10^9$-$10^{10}$ cm$^{-3}$. It was shown that non homogeneous electric field behind the IW front creates the conditions for appearance of plasma channel striations observed in the experiment. The stationary striations found in neon are more distinctive for negative polarity OED. It was shown that ionization-drift mechanism with account of metastable states existence in plasma seems more adequate for the OED striation description. This supposition allows to reasonably explain the conditions of the stationary striations existence found in the experiment.

**Introduction**

Well studied discharges of capacitive type are now widely used in various fields of plasma technology. High-frequency capacitive discharges excited by alternating voltage with frequencies in the range 1 - 100 MHz, are especially popular. Important examples of practical applications include the excitation of $CO_2$ laser based on intermediate pressure discharges (1–100 Torr), and ion treatment of materials at pressures ($10^{-3}$–1 Torr) [1, 2]. Capacitive discharge can be of the electrode type (plasma is in contact with two or one electrode) or of electrode less type (both electrodes are external). Both types of the capacitive discharge were mostly studied in plane-parallel geometry when the external electric field is uniform. The less well-known are discharges in strongly inhomogeneous electric fields involving only one electrode. The second electrode in this situation is missing and its role is played by surrounding grounded elements of the setup. In such a system, a one-electrode discharge (OED) is formed, a high-frequency form of which at atmospheric pressure and above, is also called a torch discharge [2, 3]. This discharge was named a "torch discharge" due to the shape of the plasma region, which flashed in the form of a torch around the electrode connected to a high-voltage RF generator. Now interest in this type of discharge is growing due to the possibility of its use in plasma processing of materials.

The OED studies in case of low pressure are practically not presented in recognized publications. The mechanism of such a discharge, especially at low frequencies, should radically differ from the known mechanism of ionization in a flare discharge. Such one-electrode discharge should occur in long discharge tubes with length significantly exceeding the diameter and with almost the entire external electric field concentrating near the surface of the potential electrode. This work deals with the study of mechanisms of OED ignition and glowing in a long discharge tube at gas pressures of 1-4 Torr.

A relatively large number of publications are devoted to the specific form of a capacitive discharge, outwardly similar to the OED, which was called the "discharge under unipolar gas breakdown" or UBG discharge [4–7]. This discharge is also observed in long tubes in rare gases at low pressure, however, it is generated by a single external electrode, which is a wire mesh located on the surface of the tube near one of its ends. The discharge occurs when high voltage pulses (up to 15 kV) are applied with a low repetition rate (up to 2 kHz). After application of the voltage, the plasma firstly forms in the area covered by the mesh, and then moves into the tube area in the form of a column. A distinction is made between high and low-current forms of the UBG discharge which differ from each other by not only the discharge current but also by the plasma column length at the same voltage [4]. The UBG discharge mechanism was explained on the basis of the excitation and propagation of a wave of the wall surface potential observed during the discharge.

The OED mechanism is closely related with the process of gas breakdown in a long tube. If the second electrode is connected to an external circuit, the breakdown in the tube leads to the appearance of a glow or arc discharge. In the case of a break in the low-voltage electrode circuit, only the first stage of the breakdown is realized which is generation and movement of the ionization wave (IW) [8]. The supply of voltage pulses to the potential electrode leads to periodic excitation of the IWs traveling the discharge gap. While the IW moves, the current flows in the high-voltage electrode circuit, which breaks off together with the ending of the wave movement after reaching the disconnected electrode.

The ionization wave as a breakdown mechanism (prebreakdown ionization wave) in long tubes in gases at low pressure was the subject of experimental studies in [9–12], and was also studied theoretically in [13, 14]. As a rule, these studies were concerned with the ignition of fluorescent lamps. The propagation velocity of the prebreakdown IW does not exceed $10^8$ cm/s (they are classified as slow). The IW mechanism is associated with the transfer of the high electrode potential through the discharge gap and, in general terms, can be presented as follows. When a voltage pulse is applied between the edge of the electrode and the tube wall (point of zero potential), there is the potential difference $\Delta\varphi$ between them, which trigger the avalanches developing from the initial electrons and ionization. If the electrode potential is negative, i.e. it is a cathode, then electron avalanches start near it, and their propagation is interrupted on the tube wall. The wall is charged negatively. In the case of a positive potential, the avalanches develop from random electrons present in small quantities in the gas and move in the direction of the electrode approximately in the same way as in positive corona discharge. If $\Delta\varphi$ exceeds the breakdown potential in this gap, then the concentration of charged particles increases very quickly and plasma is formed, which leads to the occurrence of the so-called initial breakdown [15].

Plasma occurred in the field of the electrode is polarized with the formation of a leading charged front, the potential of which is close to the potential of the electrode. Initial electrons begin to gain energy in the field of the front and electron avalanches generate as a result of ionization collisions. These avalanches fall into a charged front until they compensate its charge, and the space charge remaining after passing through the avalanches does not form a new front. Repetition of this process over and over again leads to the displacement of the charged front or IW. Strength of the electric field behind the IW front is much smaller than within it; the field does the work on moving the electrons to or from the electrode. Electron drift causes a current in the external circuit, which flows during the IW movement.

The above IW model requires the presence of initial electrons ahead of the front. In the case of the IW of negative polarity, the electrons are taken out by the field ahead of the front, while the high-voltage cathode being their source. It is known that negative waves appear only after a cathode spot formation [16]. In the case of positive IW there is no such permanent source of electrons and up to date there is no a common point of view on their exact nature. In [17] where the breakdown by ionization wave of a long discharge tube filled with helium was simulated at a pressure of several Torr, the high efficiency of the associative ionization mechanism in collisions of excited He* atoms with He was shown, as compared with the photoionization mechanism. The latter is often considered as the main source of the initial electrons appearing ahead the IW front. However, both mechanisms were criticized in [11] and it was shown that the most probable source of the initial electrons is the photoelectric effect at the tube wall. Partially, this point of view was confirmed in [18], where a 1–2 order reduction in the excitation time of the positive IW was observed upon irradiation of the tube walls with a source of visible light. Moreover, for waves of negative polarity this effect was not detected. In [19], a similar effect was found for plasma jets.

As the IW moves, the tube wall charges in such a way that its potential becomes close to that in the IW front. If the low-voltage electrode is not connected to the external circuit the wall charge cannot drain through the discharge column and remains on the tube wall until the voltage pulse termination. When the potential of the high-voltage electrode decreases, a potential drop occurs again between it and the wall, which leads to the repetition of the initial breakdown and the generation of another IW discharging the tube wall. Such waves were observed in [21] when a sinusoidal voltage was applied between the tube electrode and external conductive plates along the tube. In this case the occurrence of IW was observed both during the voltage growing and at its decreasing. According to the authors the second wave corresponded to the discharge of the wall. In [20], this phenomenon was discovered and studied for rectangular voltage pulses and was called the "ionization wave of return breakdown" (IWRB).

Some features of the breakdown in long tubes appear during the excitation of IW under conditions of residual gas ionization. In [23], it was shown that a decrease in the time interval between voltage pulses can result not only in the decrease in the IW induction time but also in the decrease in the breakdown total delay time. At small intervals when a high concentration of charged particles remains after the previous discharge the IW generates at a higher voltage value than at a low frequency. It was found that in this case the discharge can ignite without the IW generation but only due to ionization multiplication of residual charged particles.

All of the above mentioned phenomena are manifested to one degree or another in the mechanism of the ignition and glowing of the one-electrode discharge in long tubes, but the whole picture of the discharge physics at low pulse rate is not known. This work is aimed to the experimental study of the one-electrode discharge in neon and argon at pressures of 1–4 Torr at voltages less than 4 kV and pulse repetition rates up to 10 kHz. Under these conditions the discharge differs sufficiently in its mechanism from the known torch discharge (since it is not a RF discharge) and is not described in the literature. Due to its plasma properties it can be of practical interest for plasma-chemical and laser technologies. In this work, we suppose to give a phenomenological description of OED, obtain its main characteristics and explore its mechanism.

**Experiment**

The studies were carried out in sealed tubes 100 cm long and 1.5 cm in diameter, filled with high purity neon and argon up to pressures of 1 and 4 Torr. The tubes contained two aluminum electrodes having the form of hollow cylinders 2 cm in length and 5 mm in diameter with ceramic collars at the front edges. OED is very sensitive to the presence of any conductive objects near it. For this reason, the tube was not closed by a grounded metal screen, as it was often done in studies of the ionization waves. Instead the tube was fixed on thin dielectric stands (made of plywood) so that the distance between the tube and the nearest conductive element was at least 20 cm. Some experiments have shown that this distance is sufficient to save the discharge of perturbation.

The experimental set up layout shown in Fig. 1 differs from that in the work [24] so that the rectangular voltage pulses were applied to just only one electrode of the tube (1). Another electrode remained free (the break in the external circuit is conventionally shown via capacitor 15). The pulse repetition rate varied from 5 Hz to 10 kHz, their duration varied depending on the rate in such a way that the duty cycle of the pulses was kept constant $S = 20$, and the amplitude ranged from 1 to 3.5 kV. Both voltage polarities were investigated. The pulse front duration was determined by the response time of the switch on the field effect transistor (3) and by stray capacitances and did not exceed 100 ns, which corresponded to a voltage growth rate of $2 \cdot 10^{10}$ V/s. Ballast resistor was not installed in the circuit.

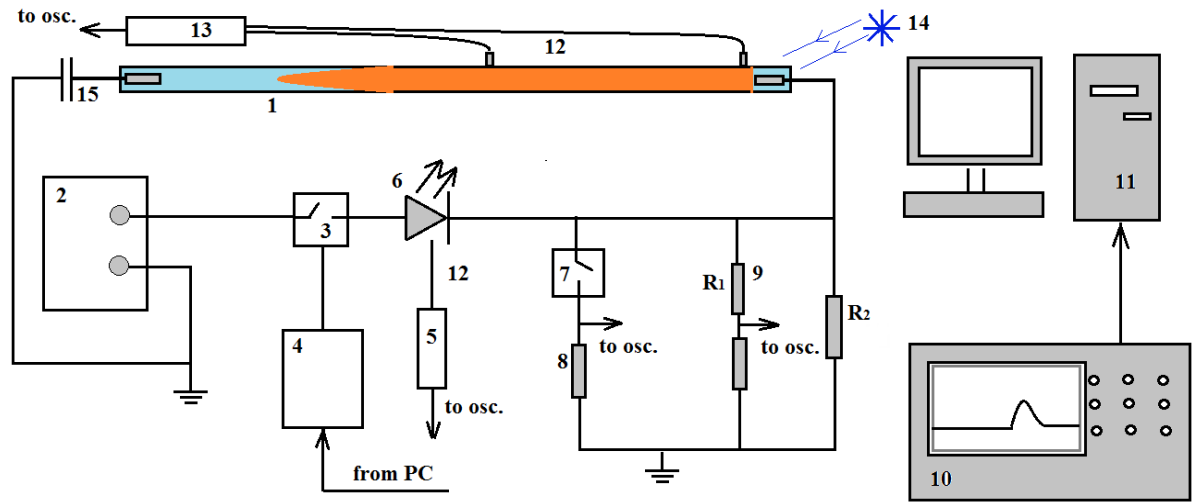

**Figure 1** The block scheme of the experimental setup. 1-discharge tube; 2- power supply; 3 - high-voltage switch; 4 - digital generator; 6 - high-voltage LED; 5, 13 - photomultipliers; 7 - switch; 8 - precision resistor (50Ω); 9 - high voltage divider; 10 - digital oscilloscope; 11 - personal computer; 12 - optical fiber; 14 - semiconductor laser (λ = 405nm); 15 - the total capacity of the tube and plasma relative to the "earth"; $R_2$ = 260 kΩ- a resistor that forms a trailing edge of the voltage pulse.

Diagnosis of OED consisted of electrical and optical measurements. The block diagram in Fig. 1 explains diagnostic system elements layout. The voltage at the electrode was measured using a high-voltage compensated divider (9) with a resistance of $R_1$ = 100MΩ, the signal from which was analyzed on Tektronix TDS 240 oscilloscope (10) with a 60 MHz pass band. The discharge current was recorded using optoelectric circuit [12], consisting of a fast high-voltage LED (6) with λ = 408 nm and a photomultiplier PMT (5) with the total response time of less than 80 ns. The LED was installed on in series with the active electrode. When a current pulse occurs in the circuit the LED generates a light flash which was detected by PMT, and then its signal is analyzed by the oscilloscope. This scheme has minimal inductance in case of short length of the connection cables. Due to distortion of the current leading edge the circuit calibration was carried out not according to the given current, but according to the charge transferred through the tube for a fixed time. To do this, the breakdown current between two electrodes of the same tube was studied. In these experiments small (50Ω) resistor (8) and opto-electrical circuit were connected in series with a low-voltage electrode. It was possible to disconnect the branch with a small resistor through the key (7). The voltage drop on the resistor is in proportion with the flowing current. Then both signals from LED-PMT circuit and the resistor were integrated over the same time interval during which the current relaxed to a value in a stationary discharge. The value of the charge $Q$, measured using a resistor was used as reference. The value of the charge $Q_d$, measured by optoelectric circuit, was compared with it. The voltage of the PMT was selected so that the device works in a linear mode and in this case the dependence $Q(Q_d)$ is linear.

Oscillographic study of the discharge radiation intensity at various points of the tube was carried out simultaneously with electrical measurements. The studies have shown that the main contribution comes from the IW front glowing. It propagates along the tube from the electrode in the form of a separate pulse. The emission of the IW was recorded using two optical fibers (12) mounted perpendicular to the tube axis and connected to one or two PMTs (13). One fiber collected the light from a point near the electrode and recorded the signal at the IW start and the second moved along the tube and transmitted the IW signal from the given point. The signals from the PMT were analyzed on an oscilloscope in particular to determine the time interval between the signals from fixed and from moving optical fibers. For a given coordinate of the fiber $x$, it was possible to determine the time of the IW movement $t$. The set of ($x$, $t$) pairs forms the $x$-$t$ diagram of the IW. An analysis of such diagrams allows us to make conclusions about the IW movement character, to trace the dynamics of its velocity as well as its reaction to experimental conditions. This method was often used in studies of high-speed ionization waves, for example in [24] and also in studies of the mutual suppression of two pre-breakdown IWs in a long tube [25].

The ionization waves parameters could not be reproduced from pulse to pulse, especially with regard to the wave speed. The reason is the finite delay time of the initial electron $\tau_d$ for the primary breakdown development. The appearance of such electron is a stochastic process, therefore, $\tau_d$ has a statistical spread. The more initial electrons are there in the gas, the shorter is this time. As a result, $\tau_d$ appears to be different from breakdown to breakdown and, consequently, the total delay time of the breakdown also changes. As a result if voltage pulse amplitude is sufficient the breakdown can occur both at the front of the pulse and at its steady state value. Another factor is the movement of the front of the IW. It is known, for example, that this process strongly depends on the initial electron concentration $n_{e0}$ near the potential electrode [8, 26].

To obtain identical ionization waves in each voltage pulse two breakdown stabilization methods were used. At a low pulse repetition rate ($f$ <50 Hz), the active electrode was irradiated by semiconductor laser (14) with λ = 405 nm. The essence of the method consists in creation of a constant value $n_{e0}$ due to electrons photodesorption from the tube wall. It is known that irradiation of a dielectric material with light having the wavelength exceeding the red boundary for a given material leads to the release of weakly bound (binding energy ≈ 2 eV) electrons from its surface [18, 22]. According to the estimates of [22], laser irradiation causes the emission of electrons with frequency of more than $10^4$ s$^{-1}$. Under the conditions of this work, the spread of the discharge ignition moments under the action of laser photodesorption decreased from 1ms up to 100ns.

The second method consisted in choosing the pulse frequency over 100 Hz. In this case, a decrease in the spread in breakdown parameters occurs due to the memory effect of the discharge gap [18, 23]. The mechanism of the effect lies in the fact that after the previous discharge, charged and excited metastable particles that maintain the value $n_{e0}$ are stored in the tube. Electrons appear when these particles collide with the metal surface of the electrode or of the wall (Auger effect) [27]. Under the influence of the memory effect, stabilization of breakdown was achieved within 1μs, which made it possible to obtain almost identical ionization waves for different voltage pulses.

**Results**

**Phenomenological image of the one-electrode discharge (OED).** OED is observed as a luminous region of the plasma occupying the entire cross section of the tube and along its length OED can occupy the entire tube or its part. The length of the plasma region is unstable and can vary depending on the conditions of the discharge exciting. In the general case OED grows from the high-voltage electrode as a front separating the plasma area from the non-ionized gas (Fig. 2).

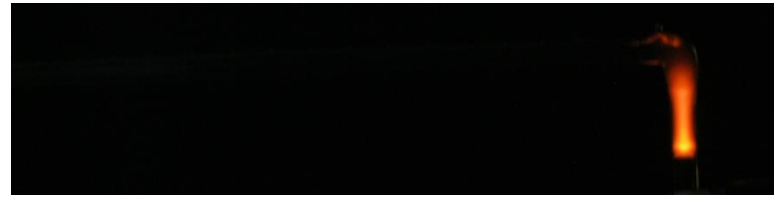

(a)

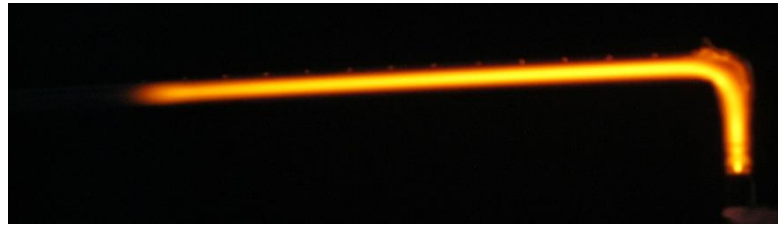

(b)

**Figure 2** One-electrode discharge in Neon 4 Torr, $f = 10$ kHz: a) minimal length; b) the discharge occupies part of the tube

The boundary of the plasma region has a conical shape and is directed away from the electrode. It propagates smoothly or jerkily with changes in voltage pulses amplitude, their frequency and polarity. In case of negative polarity the tube is filled with plasma smoothly while in case of positive voltage the OED length changes in a stepwise manner with a slight change in the discharge power parameters. The boundary of the positive polarity OED is unstable and its position varies within 1-3 cm. As a result the discharge boundary is conventionally determined as the most distant from the electrode point of OED that optical signal recorded without scatter. The length obtained $L_{cr}$ is as a rule 2–3 cm shorter than the maximum length of the plasma cone. The instability of the described boundary was not observed for discharges of negative polarity.

The length of the OED ($L$) can vary depending on the type of gas, pressure ($p$), applied voltage ($U$) and its frequency ($f$). It was found that there is a minimum value $L$, which was equal to 3-4 cm for the studied gases for positive polarity, and for negative polarity it could decrease to 2 cm. The minimum discharge length corresponds to the minimum burning voltage $U_{min}$, a slight decrease of which leads to instantaneous extinction of the OED. The results of studies of the $L$ values as a function of $U$ are presented in Fig. 3. The process of filling the tube with plasma is specific for a given gas. In neon, the sharp lengthening of discharge occurs even with a small addition to the voltage amplitude. At the same time, the growth in argon is slower and in order for the discharge to occupy the entire tube the voltages twice as high as for neon are required.

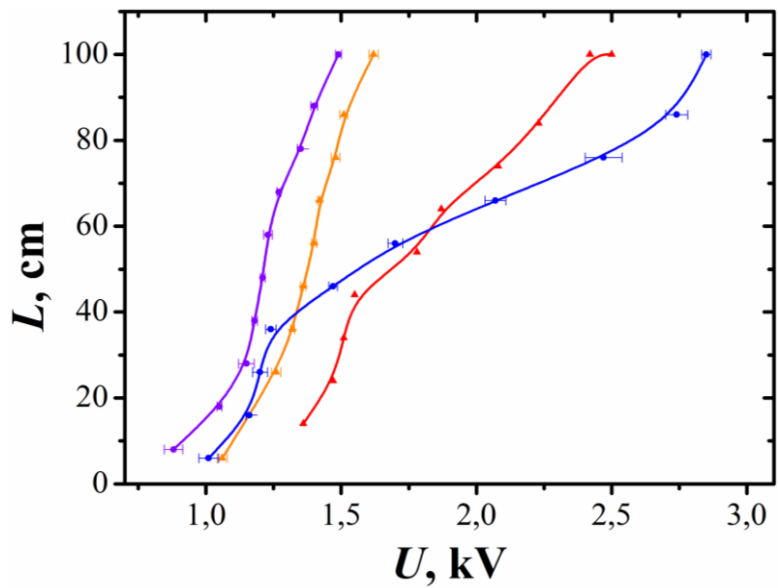

**Figure 3** Plasma length as a function of voltage amplitude, $f$ = 100Hz -▲- Argon 4 Torr, positive polarity; -●-Argon 4 Torr, negative polarity; -▲- Neon 4 Torr, positive polarity; -●- Neon 4 Torr, negative polarity

It was found that the plasma boundary shifts not only with the increase of $U$, but also with increase of pulse frequency $f$. Figure 4 shows an example of the dependence $L(f)$ for a discharge of positive polarity. At $f$ = 50 Hz and $U$ = 1.3 kV, the length of the OED is close to the minimum. As the frequency increases, $L$ grows uniformly up to 20 cm and at $f$ equal to 230-240 Hz it experiences a sharp jump to 60 cm.

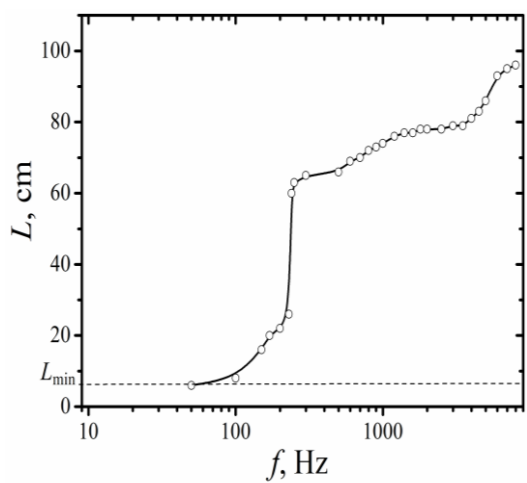

**Figure 4** Length of plasma column as a function of the pulse rate. Neon 1 Torr, $U$ = 1350 V

Further extension of the discharge proceeds smoothly with the increase of $f$. OED of negative polarity does not experience length jumps, and the $L$ values correspond to a given voltage. An increase in $f$ leads to an increase in the brightness of the plasma glow, but not in its length. It is noteworthy that the UBG discharge possesses the same properties: its length also has a jump that is observed at frequencies of ~ 100 Hz and only at positive polarity.

**Glow of the one-electrode discharge.** Glow of OED is a sequence of light pulses occurring with a frequency $f$ and having a specific structure. At $f > 20$ Hz, the eye integrates pulses over time, and the discharge glow is visually perceived as continuous, similar to a glow discharge. Figure 5 shows integral intensity oscillograms of the OED at three points of the tube: near the electrode (1), in the middle of the tube (2) and at its end. In all cases, a narrow (width $\Delta t \sim 100$ns) peak of large-amplitude radiation is observed, which corresponds to the front of the IW and transforms into a glow slowly varying with time i.e. the radiation of the plasma channel behind the front. The highest glow intensity is observed near the electrode at time $t_b$, when a discharge occurs. When the IW moves (time interval from $t_b$ to $t_{IW}$), the channel glow undergoes oscillations (striations) with a transition to a monotonic decrease. After the wave motion termination it slowly decreases and continues to be recorded up to a time $t_r$ exceeding $t_{IW}$. At points distant from the electrode only a peak of the IW front luminosity and glow of the channel are observed which breaks off after the wave motion termination.

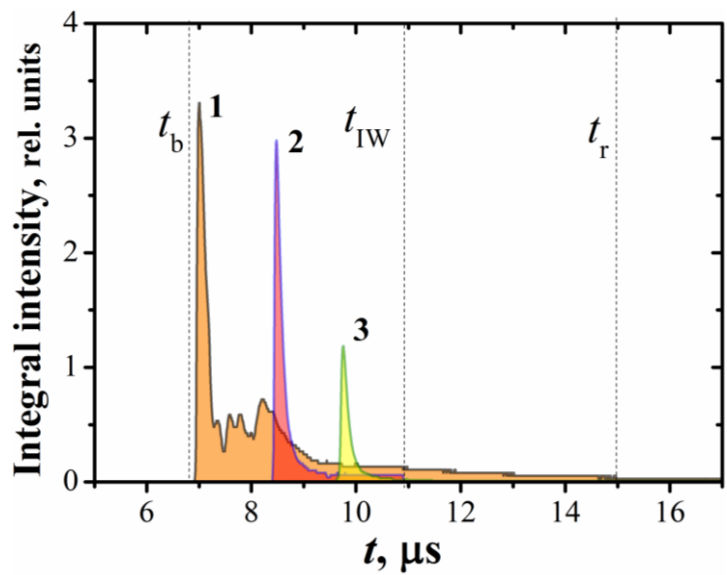

**Figure 5** Integral intensity of OED radiation in different points. Ne 4 Torr, $f = 5$ Hz, distances from the active electrode: (1) – 2 cm; (2) - 55 cm; (3) – 85 cm. Under curves areas were filled to show better its boundaries.

For most conditions studied the OED glow is diffuse but its heterogeneity along the column length is also observed which corresponds to the spatial plasma structure in the form of stationary or running striations (Fig 6). The striations glow non-uniformly: there is a brighter side which in the discharge of negative polarity is directed towards the cathode and in case of a positive polarity it is directed toward the anode. The length of the striations depends on the conditions of OED glowing and varies from a few millimeters to approximately the tube diameter. The stratification of OED is best observed in neon at a pressure of 1

Torr in a narrow range of the conditions determined by the voltage pulses amplitude, their frequency and polarity. Another important factor is the distance from the potential electrode: the larger is the distance from it, the higher is the voltage at which the striations are observed. The curves in Fig. 7 show the limits of the striations observation at a distance of 25 cm from the electrode. I - area of stationary strata existence; II - the area of violation of the stationary strata picture; III - homogeneous volume discharge. Stationary striations are clearly observed at negative polarity in the voltage range from the breakdown threshold $U_b$ up to 2.4 kV and at frequencies $f < 1.5$ kHz (Fig. 7 a). They form a distinct structure near the cathode which decays with increasing distance from it. Beyond these boundaries the regular pattern of stationary striations is violated and irregular movement occurs.

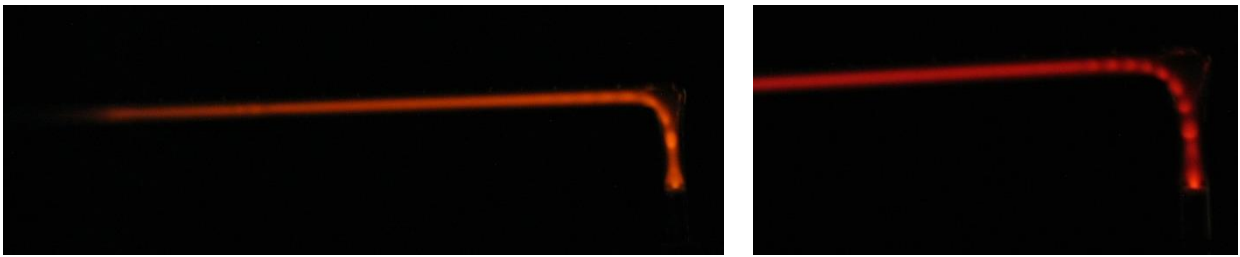

**Figure 6** Decaying stationary striations in OED of negative polarity. Neon 1 Torr, U = - 2,1 kV, pulse repetition rates: a) 1000 Hz; b) 100 Hz

With further increase in frequency or in voltage amplitude the glow becomes uniform. In case of positive polarity the region of the stationary strata existence is indicated by shading in Fig. 7b, they are observed only at high voltages and low frequencies (5Hz <f < 60Hz). However, the observation region of running striations in case of positive polarity is wider than in the negative one and strongly depends on the effect of a jump in the OED length. So at $U < 2.0$ kV, the discharge occupies a part of the tube, and its column is homogeneous.

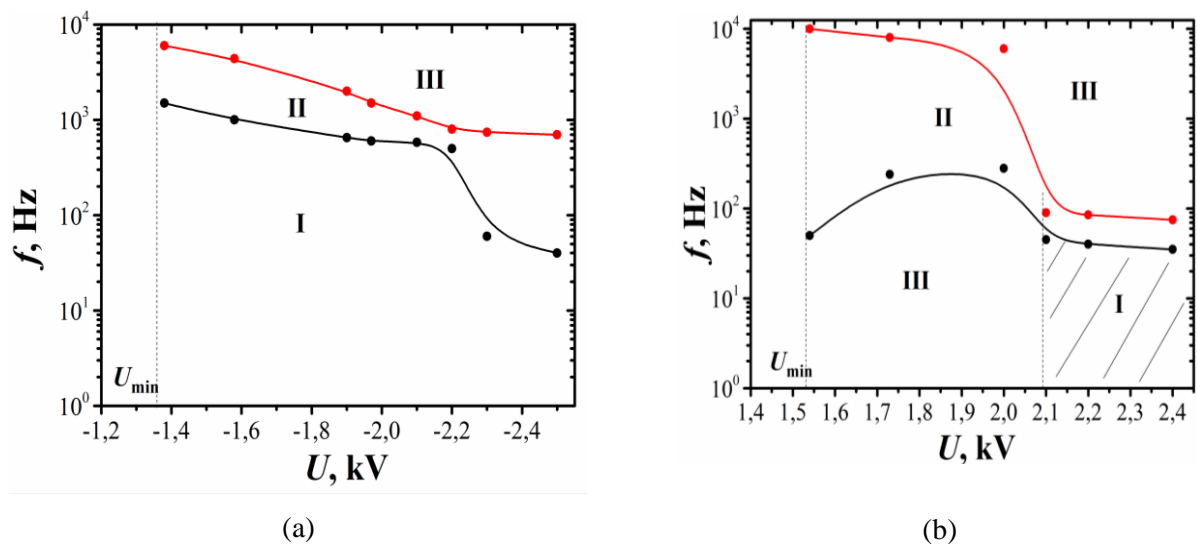

(a)                                         (b)

**Figure 7** Borders of striation in one-electrode discharge. Neon 1Torr. Red curve is the upper border of travelling striations, black curve is the upper border of stationary striations; a) negative polarity; b)

positive polarity

When the threshold frequency is exceeded the discharge lengthens abruptly and traveling striations appear which are recorded up to $f \approx 5$ kHz. At the higher voltage and low frequencies the discharge already occupies the entire tube and stationary striations can be observed but the region of the running striations existence sharply narrows. An increase in pressure up to 4 Torr causes smoothing of the striations pattern in neon. In argon no strata were observed at this pressure.

**Ignition of OED.** As was mentioned above, there is a minimum discharge length $L_{min}$ corresponding to the voltage $U_{min}$. However, the discharge is excited at a voltage $U_b > U_{min}$ with the plasma length which is several times greater than $L_{min}$. The difference in these voltages reaches 500-1000 V and depends on the frequency $f$ and the polarity.

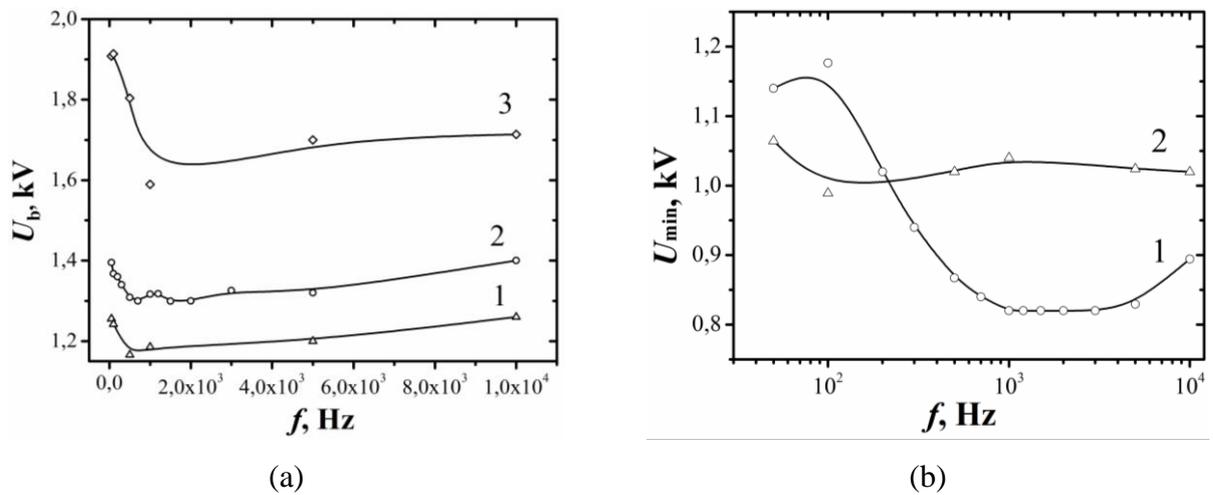

(a)          (b)

**Figure 8** a) Examples of plots for the breakdown voltage dependency on the pulse rate: 1 – Ne 1 Torr; 2 – Ne 4 Torr; 3 – Ar 4 Torr. b) Plots for the minimal voltage vs the pulse rate: 1- Ne 4 Torr; 2- Ar 4 Torr. Positive polarity.

It was found that $U_{min}(f)$ for all gases shows a decrease in the region $f < 1$ kHz, after which it reaches values that vary slightly with $f$. In neon, this effect is more expressed than in argon. Dependence $U_b(f)$ has an insignificant minimum, giving a failure of the ignition voltage at 100 - 200V, depending on the type of gas and pressure. However, the frequency dependence for $U_b$ is much weaker than for $U_{min}$. The frequency $f_{min}$, at which the voltage $U_{min}$ takes its lowest value, is $\approx 1$ kHz in neon (Fig. 8 b), and in argon $\approx 0.6$ kHz (Fig. 8 b). Ignition of OED at negative polarity (Fig. 9) is distinguished by much less voltage difference $U_{min}$ and $U_b$ (less than 200V). In neon at a frequency 100 Hz and a pressure 1 Torr this difference is completely absent, i.e. the discharge ignites and quenches at small changes in voltage, the value of $L_m$ under these conditions is the smallest (about 2 cm), and plasma glow can be observed only in the vicinity of the electrode surface. As well as in case of positive polarity, in neon the minimum ignition voltage corresponds to $f = 1$ kHz, but it is much more expressed (Fig 9 a), and the value of $U_{min}$ shows only a monotonic decrease with $f$ increasing (Fig. 9 b). The voltage $U_b$ can change significantly with a change in the humidity of the environmental air (an increase of 100V was observed) however $U_{min}$ depends only on the type of gas and on the electrode condition.

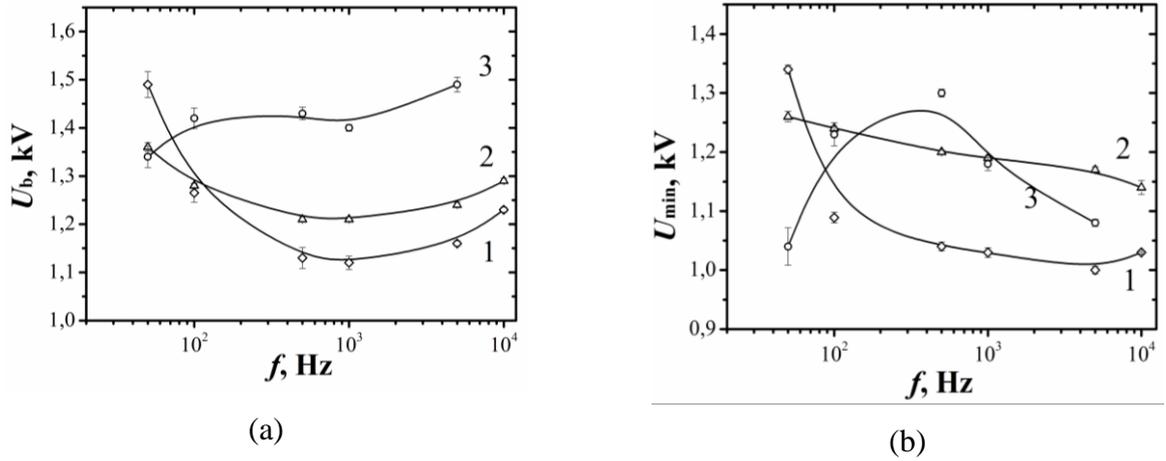

(a)   (b)

**Figure 9** a) Examples of plots for the breakdown voltage dependency on the pulse rate b) Plots for the minimal voltage dependency on the pulse rate. Negative polarity. 1 – Ne 4 Torr; 2 – Ne 1 Torr; 3 – Ar 4 Torr.

**Current of one-electrode discharge.** One of the subjects investigated was the behavior in time of the electric current through the active electrode circuit. It was found that it is basically a sequence of pulses of a specific shape that occur at the leading edge of a voltage pulse. The time course of the current depends on the voltage frequency and also changes with its polarity (the oscillograms in Fig. 10). At the leading edge of the voltage pulse a current peak is observed that occurs before the breakdown and represents the displacement current through the electrode capacitance. The further picture of the signal depends on the frequency. At $f = 5$ Hz, the peak is followed by a pause lasting from several microseconds to several milliseconds, which represents a delay in the initial breakdown. At a frequency of 100 Hz, this delay does not occur due to the memory effect. After a pause, an impulse appears with well-defined structure: a broad maximum and a small "shelf". The maximum corresponds to the initial breakdown near the active electrode after which the IW starts to move. A current corresponding to the level of the "shelf" flows to the wave front. If the IW speed doesn't change significantly during its movement, the current also changes slightly. The IW decay is simultaneous with the current pulse breaking. Increase in the frequency up to 100 Hz leads to the signal structure smoothing and to formation of a pulse with a steep edge, which is formed at the leading edge of the voltage pulse. There are also current fluctuations about the average value which remains close to a constant value during the development of IW. At negative polarity, the signal structure at low frequencies is less expressed.

We choose the average current $I_{av}$ (the amount of charge transferred through the tube per unit of time) to characterize the discharge. Since the discharge current is represented by pulses following at intervals equal to the voltage period then by the definition of the average value over the period T, we obtain:

$$I_{av} = \frac{1}{T}\int_0^T I(t)dt = \frac{Q_0}{T} = fQ_0,$$

where $Q_0$ – total charge transferred through the active electrode during the voltage pulse, $f$ – the pulse frequency. Thus, to obtain the average current it is necessary to know the magnitude of the charge transferred by the wave at a given pulse frequency. So, for example, with the increase of average current the glow intensity of the discharge column increases like it happens in DC discharges.

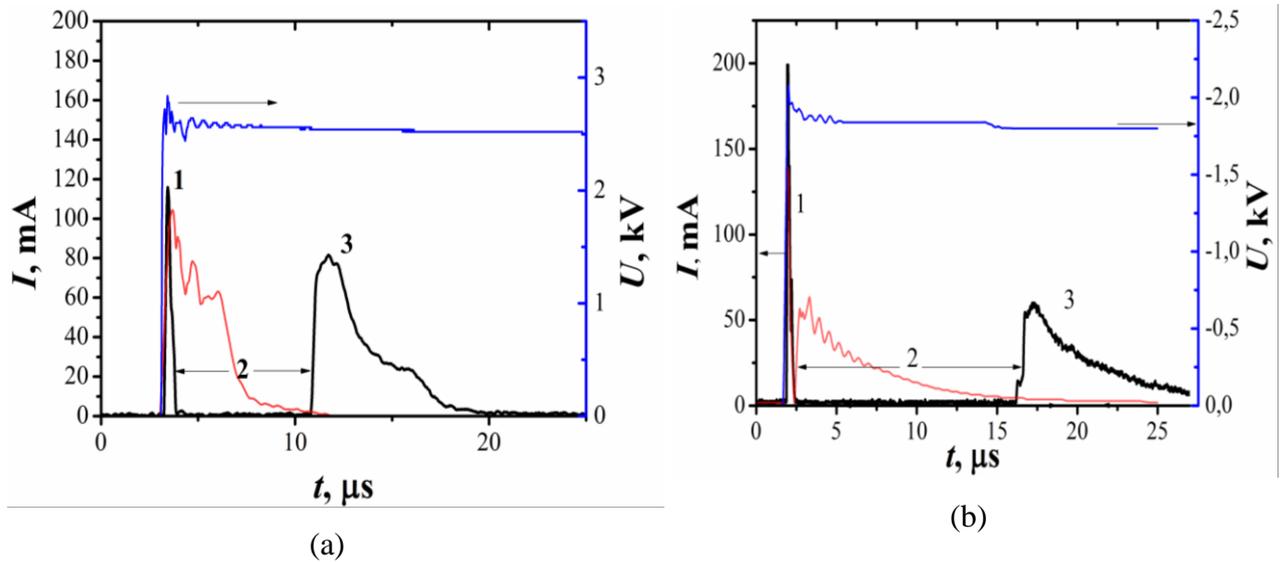

**Figure 10** Oscillograms of one-electrode discharge current at different voltage pules rates. Neon 1 Torr a) $U = 2,5$ kV blue line, $f = 5$ Hz black line, $f = 100$Hz red line; c) $U = -2$kV, blue line, $f = 5$ Hz black line, $f = 100$Hz red line. 1 – displacement current, 2 – time delay of initial breakdown, 3 – initial breakdown.

**The current-voltage characteristic of the one-electrode discharge** was obtained as the dependence of the average current on the amplitude of the voltage pulse. To plot it the dependence of the charge transferred through the active electrode on the voltage amplitude at various pulse repetition rates were made. As a result the set of the OED volt-coulomb characteristics were obtained, the examples these sets for both polarities are shown in Fig. 11.

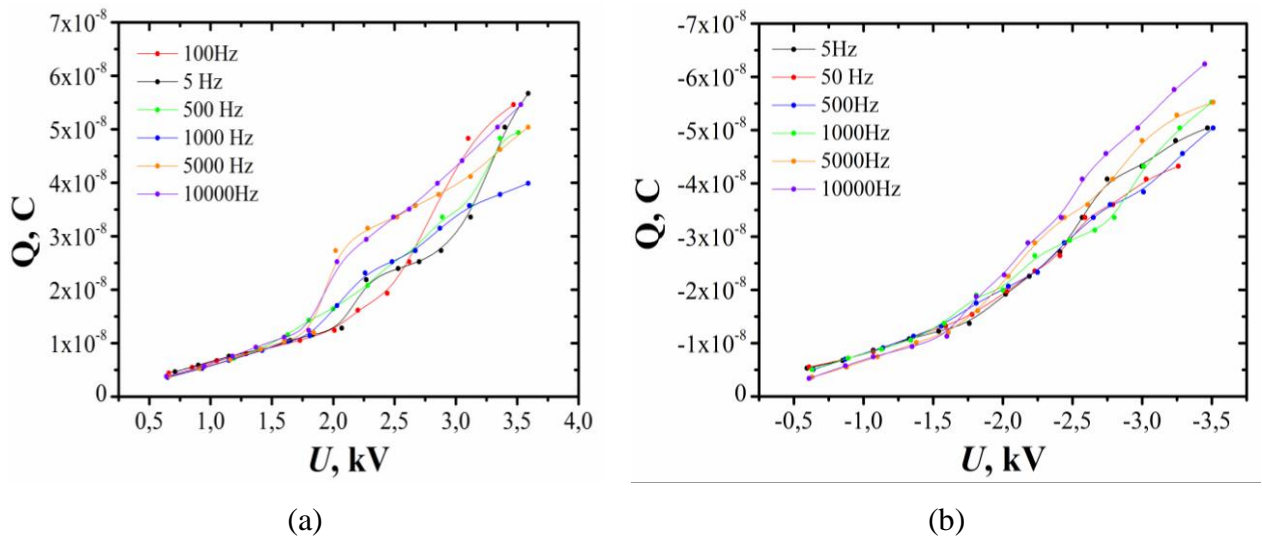

**Figure 11** Coulomb-volt curves for one-electrode discharge at different voltage pulse rates, Neon 4 Torr. a) Positive polarity of voltage; b) negative polarity of voltage.

The curves have a linear initial part up to the breakdown voltage $U_b$, after which their nonlinear growth is observed. At this moment the IW movement and plasma glow begin to be registered. In case of negative polarity the breakdown is less pronounced than in case of the positive one, however, a slightly larger amount of the transferred charge is observed. Typical example of a current-voltage characteristic (CVC) of the discharge is shown in fig. 12a. It was found that the CVC has hysteresis when the voltage increases and falls which is shown in the graph by two curves obtained for increasing and decreasing pulse amplitudes ($U$). Some specific parts of the CVC can be distinguished. 1. The beginning of the curve: $I_{av} \sim U$; 2. A sharp increase in discharge current at $U = U_b$; 3. A monotonous nonlinear increase of the curve and its reaching a slightly varying level. The graphs also show a significant increase in the discharge gap average input power after reaching the threshold $U_b$. Figure 12 b shows a graph of the discharge length as a function of increasing voltage. The hysteresis can also be noted with increasing and decreasing $U$, and then a sharp jump in length simultaneously with a jump in current. With a decrease in voltage, the current and discharge length monotonically decrease, reaching minimum values at $U = U_{min}$. In case of negative polarity the hysteresis is less manifested, or is not observed at all.

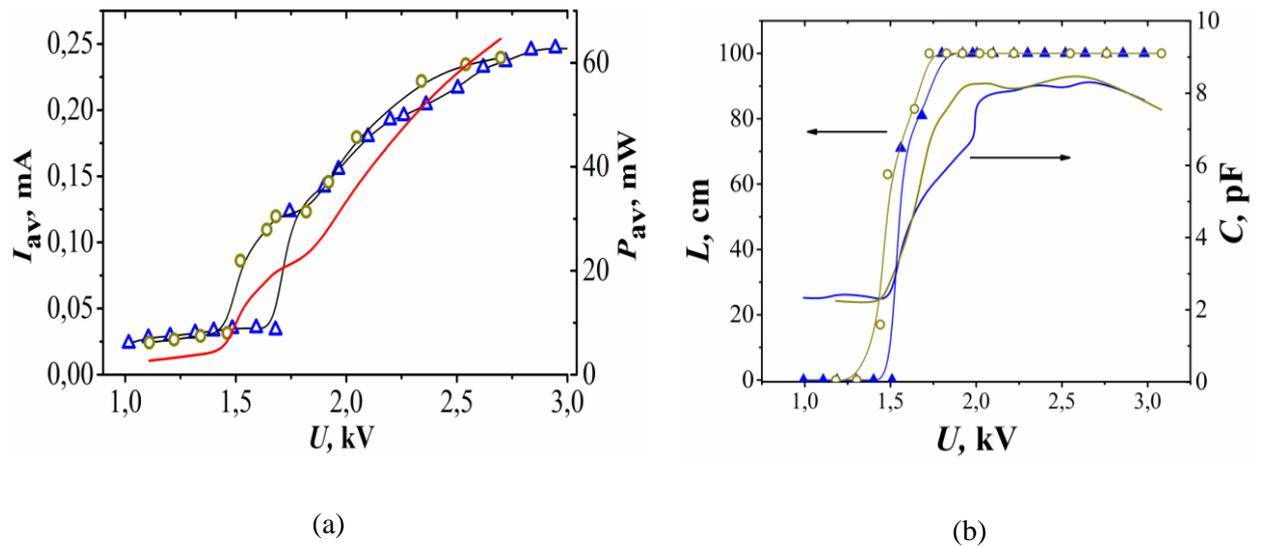

(a)　　　　　　　　　　　　　　　(b)

**Figure 12** Characteristics for one-electrode discharge of positive polarity. Neon 1 Torr, $f$ = 10Hz a) current-volt and input average power curves; b) length of the plasma column and plasma capacity. —△— increase of voltage amplitude, —O— decrease voltage, — input average power.

The structure of the CVC obtained indicates the following processes. The initial part is associated with the flow of the charging current of the electrode capacitance before ionization. In this case, the average current and voltage are proportional to each other: $\langle i \rangle = CfU$, where C is the electrode capacitance. The voltage range corresponding to this region extends from zero to $U_b$. When $U_b$ is reached, the average current abruptly increases at a constant voltage i.e. a single-electrode breakdown occurs and plasma forms in the tube volume. A further increase in voltage leads to a monotonic increase in the average current. In this case, the derivative $\frac{d\langle i \rangle}{dU}$ is higher when the discharge expands than when the tube is completely filled with plasma. We note the increasing character of the CVC as well as current-volt characteristic of single-electrode high-frequency discharges and UPG discharge. When describing the RF flare discharge [3], the presence of a capacitance between the discharge plasmoid and surrounding objects is emphasized. The AC flows through this capacitance and maintains the current in the plasmoid. In the case of the one-electrode discharge, we also deal with the plasma capacity with respect to the grounded elements of the installation, which depends on the geometric dimensions of the OED. Based formally on the definition of capacitance: $C = \langle i \rangle / fU$, it is possible to estimate the plasma capacitance of the OED for the studied voltage range. Figure 12b shows an example of such calculation together with measurements of the length of the OED. In the region $U < U_b$ corresponding to the zero discharge length, the capacitance is constant $C = 2.3$ pF - this is the capacitance of the electrode. After breakdown and the appearance of plasma the capacitance increases with voltage reaching an approximately constant value of $\approx 8$ pF after the plasma occupies the all tube.

**The propagation of the prebreakdown ionization wave** completely determines all the characteristics of a single-electrode discharge. It follows from the oscillograms of the integrated brightness of the discharge that its radiation is formed by intense luminescence of the front of the IW and a less bright luminescence of the plasma channel connecting the front and the potential electrode. The radiation from the front of the IW was recorded by a PMT in the form of a narrow peak, the detection of which determined the presence of an ionization wave at a given point in the tube. The propagation of the luminescence peak was used to study the motion of the prebreakdown IW under the conditions of the formation of one electrode discharge of a certain length. The main way to study the motion of the IW is to record *x-t* diagrams. The diagrams were recorded in different tubes at different pulse frequencies and voltage amplitudes. For these measurements, at each point of discharge *x*, up to 100 oscillograms of the optical signal of the IW were accumulated, which was compared with the signal from a point near the electrode. The difference in the time of registration of signals at point *x* ($t_x$) and near the electrode ($t_0$) gives the time the wave travels to point *x*: $\tau = t_x - t_0$. The true value of $\tau$ was taken as the sample average, which was mapped to the *x-t* diagram.

The diagrams at figure 13a correspond to the situation when positive polarity one-electrode discharge occupies a part of the tube, and its length varies depending on the voltage amplitude. The initial almost linear section in graphs 1 and 2 corresponds to the movement of the IW with an approximately constant speed. The subsequent sections of the graphs are not linear, and the nature of the curves rise corresponds to the negative value of the second derivative. The wave velocity drops rapidly (the slope of the tangent line decreases), and the diagrams transform into the horizontal lines formally corresponding to the point *x* of the IW movement termination. In reality, the wave decays when a certain minimum velocity is reached which was less than $10^5$ cm/s for conditions of the experiment. Graph 3 corresponds to the IW, which strongly attenuates from the beginning of the movement and there is no such expressed deviation of the curve. The graphs clearly illustrate the process of IW attenuation, which can be described as the motion of the IW with negative acceleration, accompanied by a decrease in the amplitude and widening of the front. If the voltage is not large enough, then the attenuation leads to the decay of the IW until it reaches the opposite end of the tube. The fact that ionization wave can decay is also known for two-electrode breakdowns [11], but this process has not been studied separately. In our experiment, the attenuation of the IW leads to the termination of the one-electrode discharge column at a certain length and plays an important role in the formation of the discharge parameters.

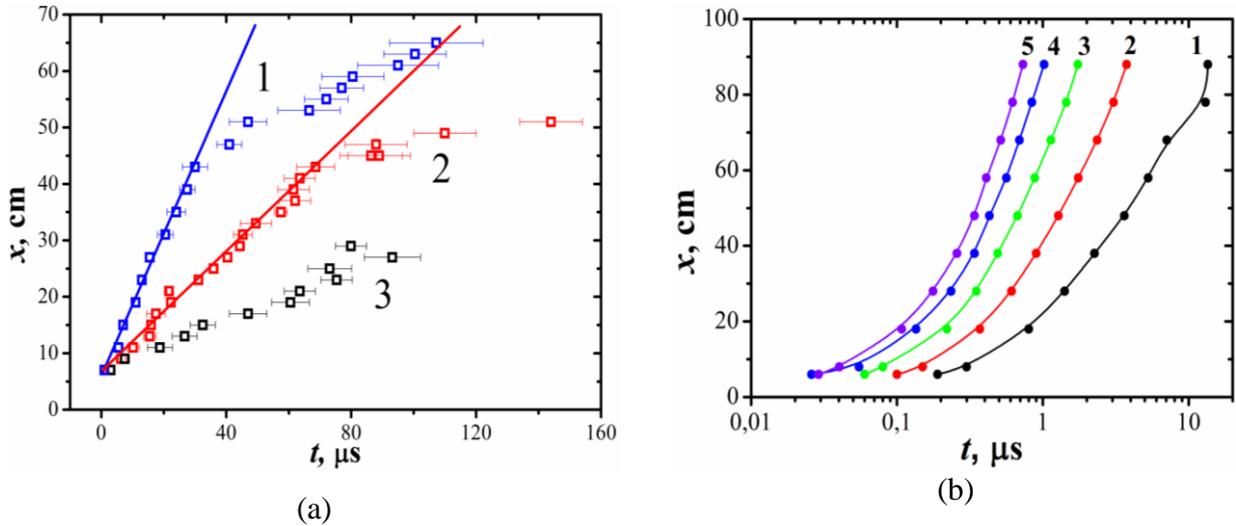

**Figure 13** *x-t* diagrams for IW of positive polarity, $f = 100$ Hz. a) one-electrode discharge occupies part of the tube. Neon, 4 Torr. **1** – $L_{cr} = 25$ cm (1260V); **2** - $L_{cr} = 50$ cm (1320V); **3** – $L_{cr} = 70$ cm (1360V). b) The discharge occupies the whole tube, Neon 1 Topp, **1** - 1,7 kV; **2** – 2,1 kV; **3** – 2.5 kV; **4** - 3,0 kV, **5** - 3,4 kV.

At the minimum breakdown delay time, the waves from different pulses travel for the same time through the same points along the axis of the tube, as if it were the same wave. But in the zone of strong attenuation, for $L > L_{cr}$, the moments of the arrival time of the IW at the point demonstrate a strong scatter which is shown in Fig.13a.

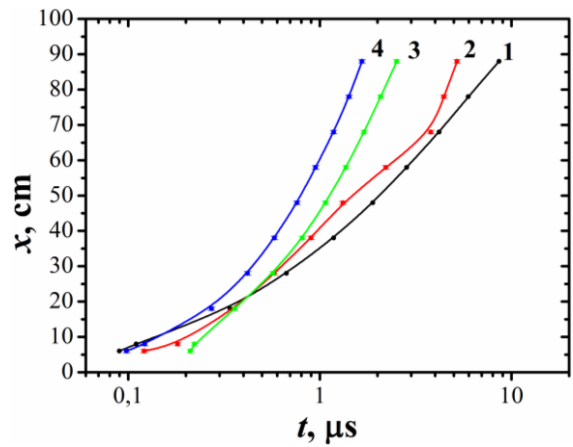

**Figure 14** *x-t* diagrams for IW in different gases, $U = 2.5$kV, $f = 100$ Hz; Argon 4 Torr 1 – negative polarity, 2 – positive polarity; Neon 4 Torr 3 - negative polarity, 4 - positive polarity.

Thus, the IW attenuation is accompanied by a violation of its propagation stability. The discharge length equal to $L_{cr}$ corresponds to the stable plasma column glow, at $L > L_{cr}$ the column luminosity is intermittent, and the length varies depending on the IW part attenuation. For negative-voltage IW such instability of the discharge boundary was not observed. However, these waves are characterized by a large attenuation and a lower propagation velocity.

If the discharge occupies the whole tube, i.e. the IW moves without decay, then the *x-t* diagrams show a monotonic increase. The graphs at Fig. 13b and Fig. 14 show *x-t* diagrams for different amplitudes of voltage pulses plotted in semi-logarithmic scale. The graphs for slow IW are more gentle and correspond to lower voltages. The observed shift of the diagrams along the abscissa axis represents a delay in the IW start. From the graphs it follows that the waves traveling large distances have a larger initial slope of the diagrams, i.e. higher initial velocity $v_0$. This parameter was separately measured at a distance of 3.5 - 4.5 cm from the high-voltage electrode as a function of its potential (Fig. 15). The graphs presented in Fig. 15 show the growth of the velocity with an increase in voltage closed to exponential for both polarities. On the graphs the dots denote experimental results, the solid lines denote approximation. At low voltage amplitudes the values of $v_0$ in different gases diverge more than at high voltages. Graphs for negative-polarity IW are characterized by a more gentle increase. For the three tubes studied the highest initial IW velocity was 150 cm/μs for the voltage amplitude 3.5 kV. During IW propagation its velocity decreases by more than an order of magnitude over the discharge column length and by 2 orders of magnitude or more at low voltages. The purpose of these measurements was to clarify the relationship between the wave velocity and the potential in its front. Since the attenuation of the IW is negligible at small distances from the electrode, the amplitude of the potential at its front coincides with the potential of the electrode. Changing the latter one can obtain a pattern of the IW speed change with a decrease in its potential due to attenuation.

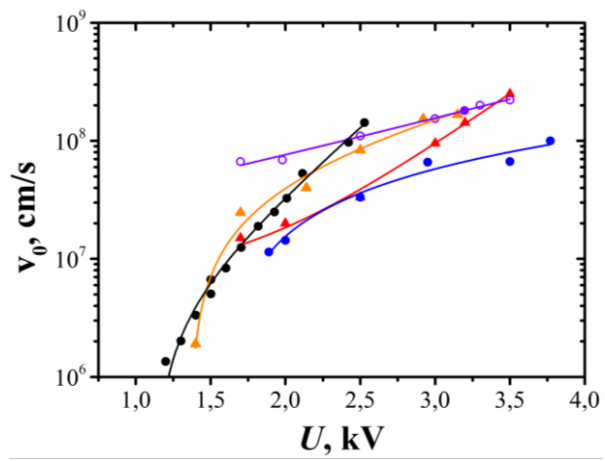

**Figure 15** initial ionization wave velocities at the beginning section of the tube, *f* = 100 Hz, -▲- Argon 4 Torr, positive polarity; -●-Argon 4 Torr, negative polarity; -▲- Neon 4 Torr, positive polarity; -○- Neon 4 Torr, negative polarity; -●- Neon 1 Torr, positive polarity.

The formation of a one-electrode discharge of a certain length is due to the specifics of the IW attenuation in a particular gas. Based on the results obtained, it is possible to propose a kinematic model of the IW motion to determine the main attenuation parameters which will be given in the Discussion section.

**Ionization wave of reverse breakdown.** In addition to the prebreakdown IW, the propagation of another ionization wave was detected at the trailing edge of the voltage pulse. According to the results of [20], the so-called reverse breakdown from the charged wall to the former potential electrode occurs. It is similar to the primary breakdown from the electrode to the wall and also leads to the formation of IW (ionization wave of reverse breakdown - IWRB), which also moves from the electrode. The direction of the current flowing behind the front of the IWRB corresponds to the polarity opposite to that of the prebreakdown IW had. Figure 16a shows examples of oscillograms of the IWRB optical signal taken at different points of the tube and demonstrating its attenuation motion. Diagrams ($x$-$t$) for different voltage polarities demonstrate the dynamics of wave motion (Fig. 16 b). At the same voltage amplitudes, the transit time over the entire tube by the IWRB is 2-3 times longer than that of the prebreakdown IW and the attenuation is higher. The experiment shows that the existence of WIRB is limited by the voltage frequency rather than by their amplitude. Table 1 shows the frequencies $f_{br}$ for the discharges in argon and neon above which the IWRB was not detected.

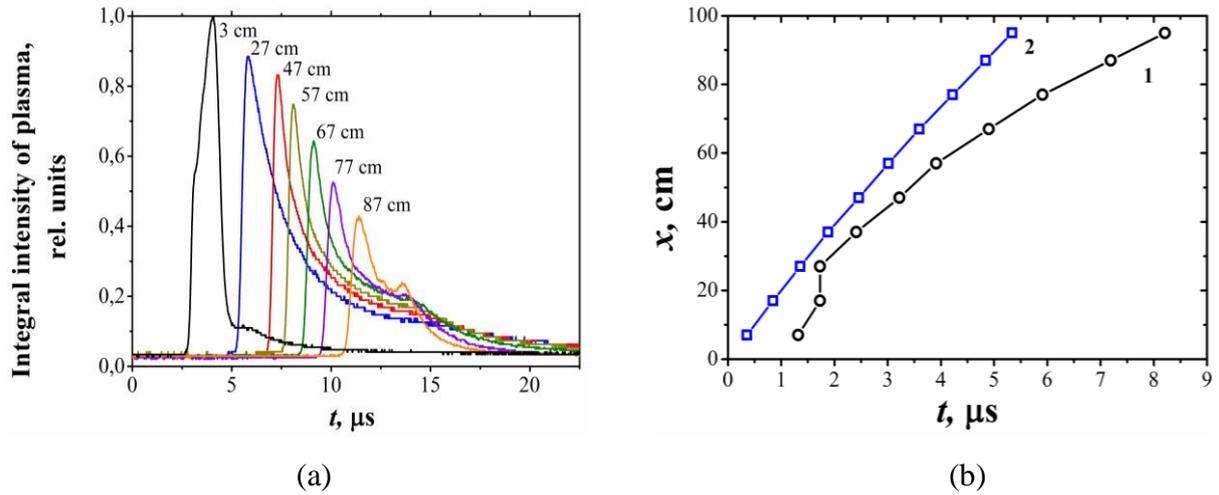

(a)          (b)

**Figure 16** Ionization wave of reverse breakdown. Neon 1 Torr, $f = 10$Hz. a) oscillograms of the optical signals from different points of OED $U = 2,5$ kV. b) $x$-$t$ diagrams for ionization wave of reverse breakdown, **1** – $U = 2,5$ kV; **2** – $U = -2,5$ kV.

**Table 1** The condition of the observation the ionization wave of the reverse breakdown

| Gas, pressure | Voltage polarity | $f_{br}$, Hz |
|---|---|---|
| Neon, 1 Torr | Positive | 150 |
| Neon, 1 Torr | Negative | 60 |
| Neon, 4 Torr | Positive | 160 |
| Neon, 4 Torr | Negative | 45 |
| Argon 4 Torr | Positive | 50 |

| Argon 4 Torr | Negative | 55 |

**Discussion**

**Mechanism of the one-electrode discharge formation in a long tube**. The results obtained indicate that the OED is formed by the periodic traveling (with frequency $f$) of the ionization waves. Moreover, the IW appears both at the leading edge of the voltage pulse (prebreakdown IW) and at the trailing edge (IW of the reverse breakdown). IW consists of a charged front carrying a high electric potential and a plasma channel through which the current flows from the front to the electrode. We can say that a high potential is transferred from the electrode to the front by the channel. If the channel conductivity were infinite, then this transmission would occur without loss, and the electric field strength behind the front would be zero. In this case, all ionization processes would occur in the front, in the region of a large potential difference [8]. However, in reality, the motion of the IW accompanied by the potential attenuation, and an electric field exists in the plasma behind the front. The lines of this field are directed along the discharge axis, approximately like in a glow discharge. In this area additional ionization of the gas and its excitation occur, which leads to the channel glow. The current in the channel flows until the IW propagates, when the motion stops, the plasma decays. The next current increase in the tube occurs simultaneously with the voltage drop at the pulse trailing edge and with the occurrence of the IWRB. Then the patterns repeat at the next discharge pulse. At some frequency of voltage, plasma no longer decays and as a result the discharge column is formed with uniform electron density in cross section and slightly varying along its length; $n_e(t)$ changes in time approximately with the voltage period. The lower is $f$ the greater is the $n_e(t)$ amplitude oscillation. The OED length is completely determined by the IW attenuation. If the wave decays at a distance from the electrode $L < L_t$ ($L_t$ is the tube length), then the discharge will occupy only a region of length $L$.

The IW movement is accompanied by the wall charging; as a result after its completion, the wall remains charged approximately up to the IW front potential and contains a total charge of ~ 10 nC, which can conserve up to several hours after the tube is quickly disconnected from the circuit [22]. The IW cannot propagate when the wall is charged because the potential difference between it and the electrode is significantly lower than $U_b$ and the initial breakdown is not possible [12, 20]. Under such conditions the IW generation would require an increase in the electrode potential up to approximately $2U_b$, which is not observed, since there is a channel for utilization of the wall charge. In our experiment there were two such channels: IWRB and charge draining into the external circuit through the plasma. At low frequencies (~ 10 Hz) the wall charge escapes mainly due to the passage of the IWRB. With increasing the frequency (see Table 1) the gas conserves the residual conductivity between voltage pulses, which leads to the charge draining through the plasma into the electrode which at this moment is connected by the switch to the "ground". For this reason IWRB degrades with increasing $f$ and is not observed at frequencies above $f_{br}$.

**Analysis of the attenuated ionization wave motion.** The analysis of the *x-t* diagrams make it possible to obtain the characteristics of the IW motion and to estimate the magnitude of the reduced electric field *E/p* and the electron concentration $n_e$ behind its front and consequently in the OED column. Let us consider the process of complete attenuation of the ionization wave along the tube, which is the most expressed when applied voltage barely exceeds the breakdown threshold. The following simplified physical picture of the attenuation can be given. The plasma channel behind the front has the finite conductivity, therefore its electrical resistance increases with IW propagation. If the channel conductivity were infinitely large, then the potential in the front $\varphi$ at any point would be equal to the electrode potential *U* and the *x-t* diagram of the IW would be a straight line. In fact, the conductivity is finite, and as the channel length increases, the voltage drop across it increases too, which leads to a decrease in the potential at the front of the IW. The lower is $\varphi$, the lower is the electric field strength ahead of the IW front and the lower is the ionization frequency. Consequently, a smaller charge enters the front and flows through the channel, which means that the current falls down and the potential drop on it increases. As a result of such feedback, $\varphi$ falls to a critical value $\varphi_{min}$, at which the ionization processes terminate and the IW decays. At a slightly higher value of the potential, the wave exists and has the lowest velocity characteristic for the IW during the OED glowing with $U_{min}$.

The empirical law that describes the decrease in the potential $\varphi$ with the increase in distance traveled by the IW *x* from the electrode is known from the results of [8, 9, 15]:

$$\varphi(x) = \varphi_0 e^{-\alpha x}, \tag{1}$$

where $\varphi_0$ is the IW potential near the electrode ($\varphi_0 \approx U$), $\alpha$ is the attenuation coefficient. The relationship between the potential in the front of the IW and its velocity was proposed to be linear in [11], but these studies were carried out for a narrow voltage range. Investigation of the IW velocity in the initial part of the tube as a function of the electrode potential shown in Fig. 15 allows for the following empirical formula:

$$v_{IW} = v_m e^{k(\varphi - \varphi_m)}, \tag{2}$$

Where $v_m$ is the IW velocity at the minimum possible value of the front potential $\varphi_m$, *k* is the coupling coefficient. At low voltages, dependence (2) becomes linear, as in [11]. Substitution of (1) in (2) gives the equation of the IW motion along the tube, in which the initial condition must be formulated as *x*(*t* = 0) = $L_m$. The choice of the minimum observable OED length as the initial condition is due to the fact that a IW forms in a certain region near the electrode, the size of which does not exceed $L_m$. The parameters $v_m$ and *k* were obtained from the approximation of the experimental data by function (2). The obtained equation was solved numerically, however, estimates showed that $\alpha \sim 10^{-3}$ cm$^{-1}$, and for most *x-t* diagrams it is permissible to use the linear approximation for equality (1). With this in mind, we can give the equation of the IW motion with slight attenuation:

$$\frac{dx}{dt} = v_0 \exp[-\alpha k \varphi_0 x], \qquad v_0 = v_m e^{k(\varphi_0 - \varphi_m)} \tag{3}$$

where $v_0$ is the initial velocity of IW at $\varphi = \varphi_0$. Its solution is represented by the analytical function (4):

$$x(t) = \frac{1}{\alpha k \varphi_0} \ln[\alpha k \varphi_0 v_0 (t - t_i) + C], \qquad C = e^{\alpha k \varphi_0 L_m} \approx 1 \tag{4}$$

in which the induction time $t_i$ of the IW determines the shift of the initial points of the *x-t* diagrams is taken into account as one of the integration constants. Approximation of the results by expression (4) demonstrates good agreement between the model and the experiment (Fig. 17 a) and allows one to determine the attenuation coefficient as an approximation parameter for various values of $\varphi_0$. If we assume that the potential of the IW at the start moment is approximately equal to the voltage at the electrode we obtain the dependencies shown in (Fig. 17 b). The following approximate formula can be proposed to describe the obtained curves under the studied conditions:

$$\alpha(\varphi_0) = \frac{A}{(\varphi_0 - \varphi_m)^n} \tag{5}$$

where the constant $A$ and exponent $n$ strongly depend on the polarity of the voltage. Table 2 shows the results of the approximation by formula (5) using the least squares method.

**Table 2**

| Gas, pressure | Voltage polarity | $\varphi_m$, V | A | n |
|---|---|---|---|---|
| Ne, 4 Torr | Positive | 1100 | 228 | 1.66 |
| Ne, 4 Torr | Negative | 1350 | 1.14 | 0.71 |
| Ar, 4 Torr | Positive | 1150 | 441 | 1.46 |
| Ar, 4 Torr | Negative | 1150 | 1.08 | 0.60 |

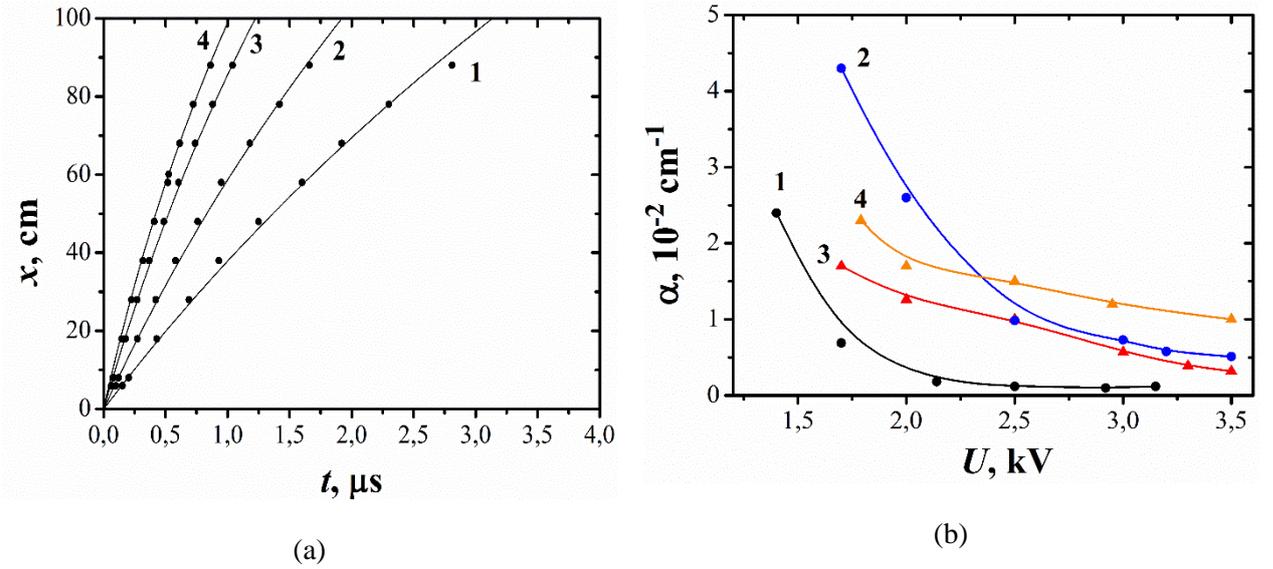

(a)

(b)

**Figure 17** a) Example of approximation the experimental results (dots) by expression (4). Ne 1Torr, positive polarity, 1 – 1.8kV; 2 – 2.5 kW; 3 – 3.0kV; 3.5 kV. b) IW attenuation coefficient as a function of applied voltage. Positive polarity: 1 – Ne 4 Torr, 2 – Ar 4 Torr; negative polarity: 3 – Ne 4 Torr, 4 – Ar 4Torr.

**Estimation of plasma parameters of OED.** Partial application of the method used in [28] for the results analysis allows to estimate the electron concentration behind the front of the IW by the magnitude of the discharge current from the relation: $I = en_e v_{dr} S$, where $n_e$ is the electron concentration, $e$ is the elementary electric charge, $S$ is the tube cross section, $v_{dr}$ is the drift electron velocity in a longitudinal electric field ($E_x$). The measured current is the conduction current through the electrode, and $v_{dr}$ was determined for the field averaged over the length of the OED. The electric field strength $E_x$ was calculated from the law of attenuation of the IW potential $\varphi(x)$ obtained by the $x$-$t$ diagram according to the assumption that it is determined by the behavior of $E_x(x)$ behind the front. Note that in case of weak attenuation the field behind the front is close to constant: $E \approx \alpha\varphi_0$; this estimation is valid at high $U$. But in the general case it is necessary to take into account the decrease in $E_x$ over the IW path. We determine the average value of the field strength over the discharge length $L$:

$$\overline{E_x} = \frac{\varphi_0}{L - L_m}(e^{-\alpha L_m} - e^{-\alpha L}) \qquad (6)$$

The electron drift velocities corresponding to the obtained mean field were taken from [29]. Since the measurements of the total charge $Q$ passing through the tube (figure 11) are more accurate than the measurements of the current, the final relation for determining the average in time electron concentration $n_e$ has the form:

$$n_e \approx \frac{Q}{eS\mu_e \overline{E}\tau}, \qquad (7)$$

where $\mu_e$ is the electron mobility, $\tau$ is the IW travel time over distance $L$. The calculation according to formula (7) for different amplitudes of voltage pulses and both polarities is presented in Fig. 18. The graphs show that at a voltage close to $U_b$, the electron concentration is ~$10^5$ cm$^{-3}$ for positive polarity and ~$10^7$ cm$^{-3}$ for negative. The difference obtained is probably due to the fact that the formation of negative IW, as is known [8], occurs after the cathode spot formation, which is accompanied by active emission of electrons from the cathode, that makes a large contribution to the detected current. As $U$ increases, the charge density increases by several orders of magnitude and reaches ~$10^{10}$ cm$^{-3}$ at $U > 3$ kV. With a positive polarity, $n_e$ grows faster than with a negative one, which leads to higher final electron concentrations.

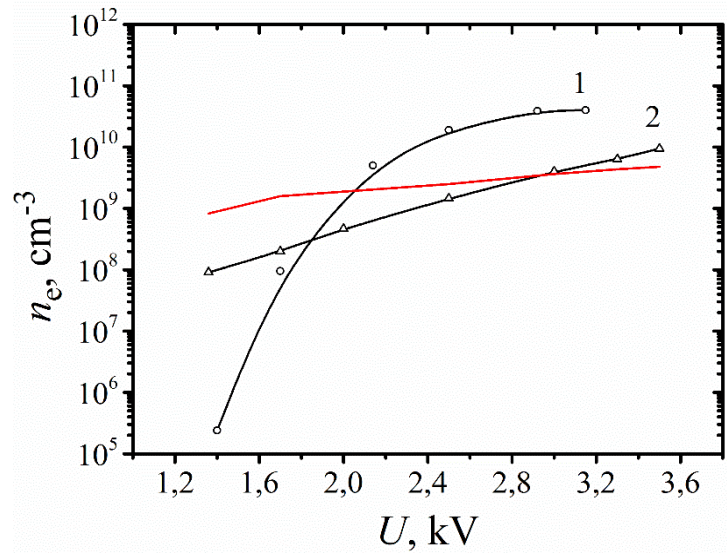

**Figure 18** estimation of average electron concentration over the OED column. Ne 4Torr, f = 100Hz, ○ - positive polarity, Δ - negative polarity.

**Conditions behind the front of the ionization wave.** Let us estimate the concentration of electrons generated by the ionization wave, as well as their lifetimes. Ionization occurs mainly in the front, if the IW did not fade, then at each point of its path it would create approximately the same number of charges, ionizing the gas uniformly. For estimation let us imagine that the front of the IW has the form of a hemisphere of radius $r_0$, coinciding with the radius of the tube, with a uniformly distributed charge over its volume. We will find the density of this charge from the assumption that on the surface of the sphere it creates a potential $\varphi$. Even if we take into account the charged channel behind the front of the IW it is still possible to give a simple estimate of the IW hemispherical head charge [30]: $Q_{IW} = 2\pi\varepsilon\varepsilon_0 r_0 \varphi$, where $\varepsilon_0 = 8{,}85 \cdot 10^{-12}$ F/m is the dielectric constant of the vacuum, $\varepsilon \approx 1$. The total charge of such a value should approximately be produced in the ionization acts ahead of the front, then merge with it and flow through the channel to the electrode, forming the OED current. The dependence $\varphi(x)$ leads to a decrease in $Q_{IW}$ along the IW path. The average value of the hemisphere charge along the IW path will be: $\overline{Q_{IW}} = \frac{2\pi\varepsilon_0 r_0 \varphi_0}{\alpha L}[1 - e^{-\alpha L}]$, which gives for $n_e \sim 10^9$ cm$^{-3}$ the values close to the electron concentrations in the negative OED (red line in figure 18). That is, taking into account all the losses we obtain that the IW must produce at least $10^9$ electrons in each cubic centimeter of gas in order to transfer high potential.

Let us consider the state of the plasma behind the front. At a pressure of 4 Torr the volume recombination is small and the main channels of electron escape from the discharge volume are drift to the electrode and diffusion. For definiteness, we consider a discharge at a voltage of 2.5 kV. According to formula (6), the average value of the reduced electric field strength is 3.4 and 3.8 V/cm Torr in argon and neon, respectively. According to the empirical dependences of $D_e/\mu_e$ on $E/N$, given in [29], the average electron temperatures for the obtained $E/N$ are 6 and 5 eV for argon and neon respectively. At the obtained charge density and average electron temperatures in the channel, as well as under the condition: $T_e > T_i$, the Debye radius is: $r_D \approx 0.05$ cm, which is much smaller than the tube radius ($r_0 = 1$ cm). Under such conditions, electron diffusion to the walls occurs in an ambipolar mode and is characterized by a diffusion coefficient $D_a$, which accepts values according to the formula: $\left(1 + \frac{T_e}{T_i}\right) D_i$, where $D_i$ is the ion diffusion coefficient. In compliance with the estimates in [31], the thermalization time of electrons for argon under conditions close to our experiment is 100 μs. For neon, this number is about 2 times less. In both cases, these values are several times greater than the duration of the current pulse and channel glow, i.e. the plasma behind the IW front is nonequilibrium. Then the characteristic time of electron escape from the plasma is equal to $\tau_d \approx 44$ μs, and $\tau_d = \Lambda^2/D_a$, where $\Lambda = \frac{r_0}{2.405}$ (for diffusion Ne$^+$ in Ne, $D_i = 20.3$ cm$^2$/s at $p = 4$ Torr [32]). For argon $\tau_d = 120$ μs (for diffusion of Ar$^+$ in Ar, $D_i = 7.4$ cm$^2$/s at $p = 4$ Torr [32]). Both values are almost an order of magnitude longer than the IW travel time.

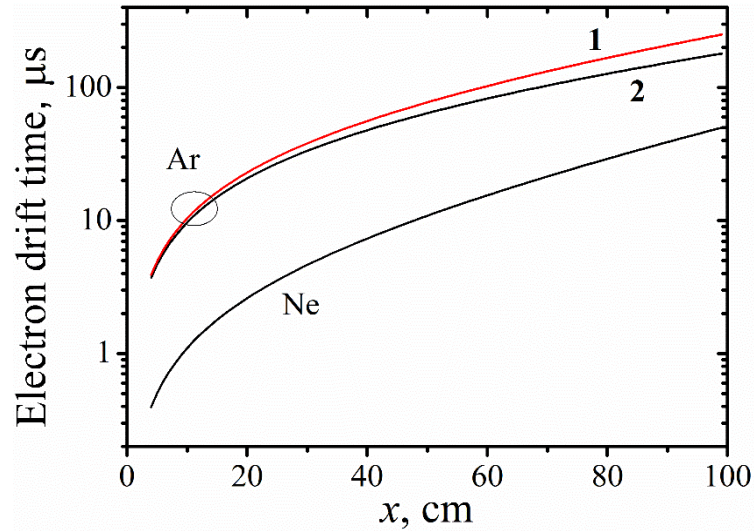

**Figure 19** Calculation of electron drift time as a function of distance from the electrode. Ar and Ne $p = 4$ Torr; 1 – $U_0 = 1.7$ kV; 2 – $U_0 = 2.5$ kV, positive polarity.

The electron drift occurs in a weak field behind the IW front, which varies exponentially according to (1). Estimates of the electron drift time from a point along the channel axis to the electrode (Fig. 19) show that it is shorter than the diffusion time, but can exceed the time of the IW motion for the most distant points of the discharge. This does not contradict the IW mechanism, since its velocity is determined by the motion of electrons in the front, and not behind it. Thus, during the existence of the discharge pulse, the charges shift to the electrode and partially disappear in the external circuit forming a conduction current. Moreover, diffusion losses can be considered insignificant.

**Decay of plasma in the channel**. Let us consider the plasma after the travel of the IW. Usually the duration of the voltage pulse in the experiment is $\tau_0 > \tau$ ($\tau$ is the IW travel time), i.e. the electrostatic field of the electrode acts on the plasma for some time after the IW decay. The external field does not act on the plasma between the pulses. The influence of the electrostatic field of the electrode reduces to maintaining a space charge near it, but the current does not occur and the electrode potential is compensated by the potential of the tube, which remains charged after the passage of the wave. Due to this, the action of the electrode electrostatic field does not lead to repeated breakdown of the gas. Since the switch maintains the constant pulses duty cycle ($S = 20$), in their duration decreases with the increased frequency, so that $\tau_0 = 1/(f \cdot S)$. As a result, at $f \sim 100$ Hz we obtained $\tau_0 \sim (4 - 10)\ \tau_d$. On the other hand, the fact that $\tau_0$ exceeds the thermalization time of electrons leads to a slowdown in ambipolar diffusion and an increase in $\tau_d$. If we assume that the diffusion lifetime is minimal until the temperature of electrons and ions is equalized, (corresponds to the largest value of $D_a$), then the plasma density will decrease by no more than an order of magnitude during the thermalization. Under the conditions of equal electronic and ionic temperatures the ambipolar diffusion slows down ($D_a = 2D_i$) and the value of $\tau_d$ exceeds the pulse duration. As a result, the plasma density during the $\tau_0$ decreases by no more than 1 - 2 orders of magnitude, and the IWRB at the trailing edge of the voltage pulse occurs under gas preionization. If the latter is such that the gas conductivity achieved in the discharge channel is preserved, then the IWRB may not occur, especially in case of a low-slope trailing edge of the voltage pulse (see Table 1). Indeed, the diffusion lifetime of electrons in argon is almost three times longer than in neon; therefore, the concentration of electrons critical for the IWRB development in neon will be reached at $f$ three times higher than in argon. This is precisely the result obtained in the case of the positive polarity OED. At negative polarity, the critical frequency for the IWRB generation remains the same in argon, while in neon it becomes three times lower.

In case the voltage frequency exceeds the critical one when the IWRB is not formed, connection of the electrode to the ground leads to the flowing of charge from the tube wall through the remaining plasma. This process is followed by the plasma decay without the external fields and charges on the wall and electrode. Let us estimate the electron escape time under the above conditions. At high frequencies and a short duration of the voltage pulse when the charge leaves the walls, the electron concentration differs slightly from the values in the discharge, i.e. $n_e \sim 10^9 - 10^{10}$ cm$^{-3}$; plasma decay occurs due to ambipolar diffusion. For $f < 1$ kHz, the time from the moment of the IW motion termination up to the start of the next pulse significantly exceeds the electron thermalization time. For these frequencies, we can assume that the ambipolar diffusion coefficient is determined by the expression: $D_a = 2D_i$, and the characteristic electron escape time in neon at $p = 4$ Torr is $t_{dif} = 4.2$ ms, and in argon at the same pressure

$t_{dif}$ = 11.7 ms. The plasma lifetime can be estimated as 10 $t_{dif}$, which indicates that deionization of neon does not occur already at $f > 50$ Hz, and that of argon at $f > 10$ Hz. Starting from these frequencies, the passage of IW does not create plasma, but only restores the charge particles concentration. At a high frequency this effect will lead to accumulation of charges in the tube and to smoothing of $n_e$ and $n_i$ modulation.

**OED column striations.** The above estimates of the electrons concentration in the plasma of the OED were made under the assumption that its distribution along the discharge is uniform. At low frequencies and voltage amplitudes, this approximation is unjust. Under these conditions, the discharge length is smaller than the length of the tube, and difference in the potential between the electrode and the plasma boundary is large, which is manifested in strong attenuation of the IW. As a result of this, the ionization becomes strongly non uniform along the discharge column. In the region of the initial breakdown, the electron concentration is high and decreases towards the boundary of the OED. On the other hand, a space charge layer is formed near the electrode, in the manner similar to the one occurring in a single-electrode barrier discharge [1, 2]. The resulting inhomogeneity of the charge concentration leads to inhomogeneity in the distribution of the electric field along the discharge column. Near the electrode, a zone of increased electric field strength appears, which can cause the appearance of fast electrons. The fast electrons beam can ionize the gas outside this zone, in a way similar to the electrons that have passed the cathode layer in a glow discharge. It is known that in helium the relaxation of such beams leads to damped oscillations of the electron density and electric field strength upon transition from the Faraday dark space to the positive column. These fluctuations are observed in the form of one or two stationary striations. The attenuation decreases when the stepwise ionization exists in the presence of metastable atoms, and several striations can be seen [33]. A similar situation can arise behind the front of a rapidly attenuated IW in response of significant inhomogeneity of the electric field and electron density.

For the stationary striations observed in our experiment, we can propose an ionization – drift formation mechanism [34], which was considered for striations in a pulsed nanosecond discharge in inert gases [35]. This mechanism manifests itself at medium gas pressures and low voltage amplitudes. It is shown that under these conditions the electron energy balance is determined mainly by inelastic collisions, and the length of the electron energy relaxation with respect to inelastic collisions is much less than the energy relaxation length with respect to elastic collisions ($\lambda_\varepsilon$). As a result, electrons accelerated by an external field gain enough energy for ionization over a length: $l \geq \varepsilon_{ex}/e\bar{E}$, (where $\varepsilon_{ex}$ is the excitation threshold), which they subsequently lose due to inelastic impacts with formation of a group of slow electrons. These slow electrons are again accelerated by an external field until inelastic collisions with the loss of the energy $\varepsilon_{ex}$, etc. The length $l$ determines the scale of the inhomogeneity of the electric field in the plasma, i.e. the length of the striations, and $\lambda_\varepsilon \gg l$ [34, 36]. Let us confirm the possibility of such a mechanism for the formation of striations in the OED by the following estimates.

For definiteness let us assume that the electric field has an exponentially declining profile $E(x) \approx \alpha\varphi_0 e^{-\alpha x}$, which determines the decrease in the IW potential, while the attenuation coefficient is sufficiently large and the wave decays in the middle of the tube. The relaxation length of electron energy due to elastic collisions is $\lambda_\varepsilon = \delta^{-\frac{1}{2}}\lambda_e$, where $\delta = 2m/M$ is the energy transfer coefficient ($\delta = 5.52 \cdot 10^{-5}$) and $\lambda_e$ is the mean free path of an electron with respect to elastic collisions. At pressure 4 Torr, $\lambda_e \approx 0.025$ cm, and the electrons reach their average energy at the distance $\lambda_\varepsilon \approx 5$ cm, which exceeds the observed sizes of the striations (1 cm). Therefore, if we take into account the high electric field strength near the electrode, we will meet the conditions allowing the formation of a beam of fast electrons that gain the energy equal to the ionization threshold over a length shorter than $\lambda_\varepsilon$. The mean free path of an electron associated with ionization collisions can be estimated by the formula $\lambda_{\varepsilon e} = (n_a \sigma_{ex})^{-1}$ (where $n_a$ is the concentration of atoms), and the total cross section of excitation near the threshold can be approximated by the relation: $\sigma_{ex} \approx C(\varepsilon - \varepsilon_{ex})$, where $\varepsilon$ is the electron energy, the constant $C$ for neon: $1.5 \cdot 10^{-18}$ cm$^2$/eV; $\varepsilon_{ex} = 16.8$ eV [1]. The cross section depends on the electron energy, which in turn is determined by the electric field strength varying along the discharge length; therefore, the quantity $\lambda_{\varepsilon e}$ will also change with the distance $x$ from the electrode. Let us evaluate this change, assuming that the electron gains energy $\varepsilon$ along the length $\lambda_{\varepsilon e}$ and $\varepsilon = eE(x)\lambda_{\varepsilon e}$. In that case:

$$\lambda_{\varepsilon e} = \frac{1}{2eE}\left(\varepsilon_{ex} + \sqrt{\varepsilon_{ex}^2 + \frac{4eE}{n_a C}}\right), \tag{8}$$

The results of calculations by formula (8) for a negative-polarity OED are shown in Fig. 20 for $U = 1.7$ kV. The graphs show that on the first 15 cm of the OED length, where stationary striations are recorded, the condition $\lambda_\varepsilon \gg \lambda_{\varepsilon e}$ is satisfied. The values $\lambda_{\varepsilon e} \approx 1$cm, which are close to the observed sizes of the dark regions between the striations Thus, the conditions for the proposed mechanism of striations are fulfilled.

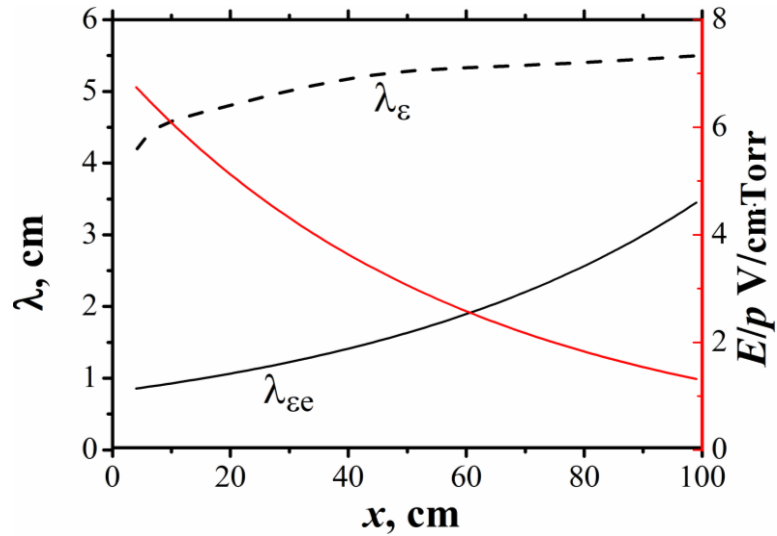

**Figure 20** The calculated values of the relaxation lengths of the electron energies with respect to elastic ($\lambda_\varepsilon$) and inelastic collisions ($\lambda_{\varepsilon e}$), as well as the reduced electric field as functions of the distance from the active electrode. Neon 4 Torr, $U = 1.7$ kV, negative polarity.

The dependence of striations existence on the frequency and voltage amplitude is apparently related to the ionization conditions in the plasma channel. An important factor is the presence of metastable atoms produced in the IW front. In the region of high electric field strength of the front, not only ionization of atoms occurs, but also the filling of metastable states [37] takes place.

The lower bound on the lifetime of the metastable state Ne$^*$($^3P_2$) in the channel behind the wave front is obtained by neglecting the processes of the state filling in the discharge. The diffusion of metastable atoms to the wall with subsequent quenching on it occurs over the time $\tau_m$ = 4.6 ms, with $\tau_m = \Lambda^2/D_m$, where the diffusion coefficient $D_m$ = 23.9 cm$^2$/s is calculated according to the data from [38] for pressure 4 Torr. At $n_e \sim 10^8$ cm$^{-3}$, which is characteristic for the striations observation conditions, the contribution of diffusion to the metastable states quenching dominates over the contribution of superelastic collisions with electrons (the process constant is $k_e$ = 4·10$^{-7}$cm$^3$/s [38, 39]). However, at such pressure, the significant contribution is also made by quenching mechanism with neon atoms in the ground state in two and three particle collisions. The corresponding process constants are: $k_2$ = 3.4·10$^{-15}$cm$^3$/s and $k_3$ = 5·10$^{-32}$cm$^6$/s [38, 39]. By taking these processes into account the characteristic lifetime of a metastable atom can be reduced to about 0.7 ms, which can be taken as the lower boundary of this value. This characteristic lifetime can be increased by taking into account the processes of filling of a metastable state (for example, by electron impact from the ground state). The obtained characteristic lifetime is two orders of magnitude longer than the time of the IW motion, therefore, the concentration of Ne$^*$ in the plasma behind its front will change insignificantly during the channel existence and remain approximately at the level reached in the wave front.

The presence of metastable particles leads to an additional contribution to the ionization in the channel due to stepwise processes, which can occur at a lower electric field strength. As noted above, stepwise ionization from metastable levels leads to decrease in the attenuation of stationary striations arising behind the cathode layer of a glow discharge in inert gases [33]. Thus, the presence of metastable atoms creates favorable conditions for the development of stationary striations at a considerable distance from the electrode, which is observed in the OED. In fact, the obtained lifetime is close to the inverse voltage frequency at the boundary of the stationary striations existence in the diagram of Fig. 7a. These two quantities can be brought in even closer proximity by taking into account the processes of filling of metastable states in the channel and their accumulation with an increase in the pulse frequency. On the other hand, the accumulation of metastable atoms will lead to an increase of stepwise processes contribution to the ionization. Consequently, as $f$ increases, the $\lambda_{\varepsilon i}$ 1will decrease, and the stationary striations pattern will smooth. So on the one hand, the presence of metastable atoms in the channel enhances the inhomogeneity of the plasma but on the other hand, their accumulation leads to easier ionization at low field strengths and evens out the concentration of charged particles along the discharge column. These processes can limit the existence of stationary striations by the pulse frequency and determine the boundaries in the diagrams in Fig. 7.

**Conclusions**

The purpose of this study was to research low-frequency one-electrode discharge in long discharge tubes in inert gases, argon and neon at pressures of 1–4 Torr. We investigated the conditions of its ignition and suppression, the main electrical characteristics, as well as the dependence of the length of the plasma column on the amplitude and frequency of the voltage pulses. We also obtained kinematic characteristics of the prebreakdown ionization wave, which is the plasma generation and maintenance mechanism. Typical concentrations of charged particles in a nonequilibrium plasma of OED are $10^8$-$10^{10}$ cm$^{-3}$ and are not the same for discharges of different polarity. It was found that the length of the OED column is determined by the attenuation of the IW during its propagation from the active electrode and exceeds a certain minimum distance from the electrode needed for the wave formation. It is shown that the observed luminous plasma column of the OED represents a repeatedly flashing channel behind the front of a traveling ionization wave. The attenuation of the IW amplitude leads to the longitudinal electric field in the channel which is sufficient for excitation and ionization of the gas atoms. In case of strong attenuation the inhomogeneity of the electric field and the concentration of charged particles along the discharge column can be large, which leads to the development of instabilities in the form of stationary striations. The boundaries of the striations existence are determined by threshold values of voltage and its frequency. It was shown that striations are most easily observed near the breakdown threshold at frequencies shorter than the inverse lifetime of metastable atoms. Unlike a stationary glow discharge, the latter are efficiently produced by the ionization wave and remain in the OED plasma. At a sufficiently high frequency of voltage pulses metastable atoms and charged particles accumulate in the discharge volume which leads to uniformity of the plasma column. In conclusion, we note that, despite the fact that the high-frequency case of the one-electrode discharge in a long tube has been studied previously (torch discharge at low pressure), no description of its low-frequency version can be found in widely known works. However, the mechanism of plasma generation by an ionization wave provides high densities of charged and excited particles, including metastable ones, which opens up prospects for various plasma-chemical applications.

**Acknowledgments**

The authors express their deep gratitude to Professor Yu. Z. Ionikh. for careful reading of the manuscript, stimulating discussions and important remarks.

The reported study was funded by RFBR according to the research project № **18-32-00223**